\documentclass[%
 aip,%
 cp,%
 amsmath,amssymb,%
 reprint,%
 groupedaddress,
]{revtex4-1}

\usepackage{graphicx}
\usepackage{dcolumn}
\usepackage{bm}

\begin{document}

\preprint{AIP/123-QED}

\title{Identification of the low-energy excitations in a quantum critical system}

\author{Tom Heitmann and Jagat Lamsal}
\affiliation{Missouri Research Reactor, University of Missouri, Columbia, MO 65211, USA}%
\author{Shannon Watson, Ross Erwin, Wangchun Chen$^{\text{a)}}$, and Yang Zhao}
  \altaffiliation[Also at ]{Department of Materials Science and Engineering, University of Maryland, College Park, MD, USA.}
\affiliation{National Institute of Standards and Technology, Gaithersburg, MD 20899, USA}%

\author{Wouter Montfrooij}
 \email{montfrooijw@missouri.edu.}
\affiliation{Department of Physics, University of Missouri, Columbia, MO 65211, USA}%

\date{\today}

\begin{abstract}
We have identified low-energy magnetic excitations in a doped quantum critical system by means of polarized neutron scattering experiments. The presence of these excitations could explain why Ce(Fe$_{0.76}$Ru$_{0.24}$)$_2$Ge$_2$ displays dynamical scaling in the absence of local critical behavior or long-range spin-density wave criticality. The low-energy excitations are associated with the reorientations of the superspins of fully ordered, isolated magnetic clusters that form spontaneously upon lowering the temperature. The system houses both frozen clusters and dynamic clusters, as predicted by Hoyos and Vojta [Phys. Rev. B 74, 140401 (R) (2006)].
\end{abstract}

\maketitle

\section{\label{sec:level1}Introduction}

Quantum critical systems are systems that exhibit a competition between the tendency for magnetic moments to order upon cooling and the tendency of those moments to be shielded by the conduction electrons through the Kondo shielding mechanism. Through the application of hydrostatic pressure or chemical pressure (doping), the phase transition to a long-range ordered state can be suppressed to zero Kelvin; when this happens, such systems have been observed to display non-Fermi liquid behavior over a range of temperatures\cite{stewart}, and some such systems display dynamical scaling. When a quantum critical system displays dynamical scaling, then the response is no longer a function of the energy $E$ transferred to the system at temperature $T$, but of the ratio $E/T$ only. This scaling was first observed in UCu$_4$Pd\cite{meigan}, and has later been identified in other quantum critical systems\cite{stewart,schroder,wouterprl}, also as a function of applied magnetic field\cite{schroder} ($H/T$-scaling).\\
In order for a system to display dynamical scaling, low-energy degrees of freedom must be available for excitations to couple to and relax back to equilibrium. If this is the case, then the $E/T$-scaling is similar to the emergence of scaling behavior in high-energy physics where we also have the situation that $T \gg E$. The heavily doped quantum critical system Ce(Fe$_{0.76}$Ru$_{0.24}$)$_2$Ge$_2$ represents a very interesting case in the sense that this system displays $E/T$-scaling\cite{wouterprl}, but two plausible scenarios\cite{hertz,millis,si} for the origin of low-energy excitations have been ruled out. Experiments on poly-crystalline Ce(Fe$_{0.76}$Ru$_{0.24}$)$_2$Ge$_2$ have already shown\cite{wouterprl} that there is no divergence of the local susceptibility in this system (that would reflect the emergence of local moments at the expense of Kondo shielded moments\cite{schroder,si}), and that the dynamic susceptibility has an energy dependence that  cannot be described by Lorentzian lineshapes, excluding a scenario\cite{hertz,millis} where the dynamical scaling is ascribed to fluctuations associated with a transition to a long-range spin-density wave (SDW) ordered state.\\
A solution to this observed $E/T$-scaling conundrum could be the presence of clusters.
Ce(Fe$_{0.76}$Ru$_{0.24}$)$_2$Ge$_2$ has been shown\cite{wouterprl} to form isolated, magnetically ordered clusters upon cooling down. When the temperature is lowered, the magnetic moments on the Ce ions get shielded by the conduction electrons through the Kondo shielding mechanism. However, since the system has one Ru ion for every three Fe ions (randomly distributed), the ensuing distribution in interionic distances has been shown to result in a distribution of Kondo shielding temperatures. As a result, some moments get shielded before others, and the system displays percolative behavior\cite{stauffer} upon cooling. The clusters that peel of the lattice spanning, infinite cluster are subject to quantum mechanical finite-size effects\cite{heitmannconf}; as a result, the moments of the ions that constitute the cluster align with their neighbors, giving rise to fully ordered magnetic clusters with Ising symmetry. The superspin of such a cluster is free to pick a direction along the positive or negative z-direction (c-axis). As such, these clusters have one degree of freedom left, namely the direction of the superspin along the Ising axis. Reorientations of the superspin could provide the sought-after low-energy excitations that this system must be capable of in order to explain the observed $E/T$-scaling behavior.\\
The formation of these isolated clusters has been observed directly by means of neutron scattering experiments\cite{wouterprb}, and the unusual behavior of the low temperature specific heat was shown\cite{gaddy} to be consistent with the entropy shedding expected for clusters becoming isolated and forced to order magnetically. Monte Carlo computer simulations for the relevant percolative system\cite{heitmannconf,heitmannonline} have shown that the emergence of magnetic clusters should play a dominant role in the low temperature behavior of heavily-doped quantum critical systems, while a comparison to a purely classical system\cite{heitmannprb} that houses magnetic clusters has demonstrated that the observed dynamical scaling in quantum critical systems is likely to be the consequence of the presence of isolated clusters. In this paper, we demonstrate that superspin reorientations of isolated clusters do indeed occur in Ce(Fe$_{0.76}$Ru$_{0.24}$)$_2$Ge$_2$, and that their flipping rates are qualitatively given by the predictions of Hoyos and Vojta\cite{vojta} for isolated Ising clusters of varying sizes.
\section{Results and Discussion}
The predicted\cite{vojta} dynamic behavior of isolated clusters in an Ising system with dissipation depends on the size of the clusters: the rate at which the superspin of a cluster can flip depends on the size of the cluster, and above a critical cluster size the spins are frozen in. We argue that quantum critical Ce(Fe$_{0.76}$Ru$_{0.24}$)$_2$Ge$_2$ is such a system.\\
\begin{figure}
\begin{center}
\includegraphics*[viewport=30 75 480 630,width=70mm,clip]{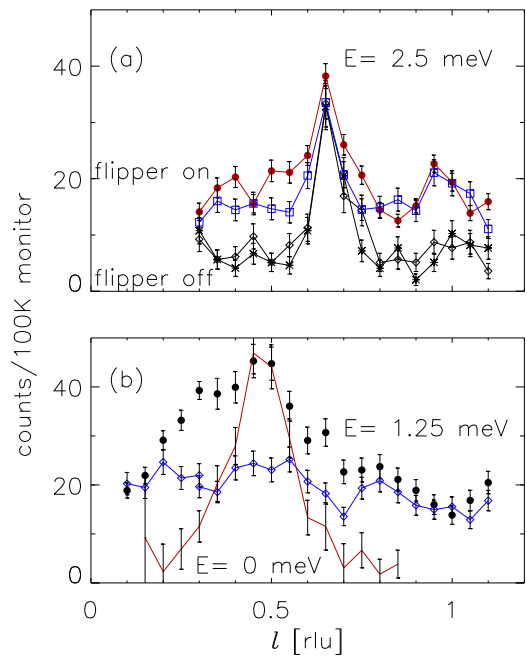}
\end{center}
\caption{a) The top panel displays polarized scattering data measured along [1,1,$l$] in reciprocal lattice units (rlu) for energy transfer $E$= 2.5 meV, corresponding to excitations with wave number $2\pi l/c$, with $c$ the length of the c-axis. Both the flipper on (magnetic, top two sets) and flipper off (nuclear, bottom two data sets) are shown for $T$= 1.56 K and 50 K (circles: 1.56 K, magnetic; squares: 50 K, magnetic; stars: 1.56 K nuclear; diamonds: 50 K nuclear). The magnetic data shows that there is broad scattering present at 50 K, and that this scattering increases upon cooling to 5 K over a broad $l$-range. The feature at $l$= 0.65 is a spurion. b) The bottom panel displays the magnetic, flipper on data for $E$= 1.25 meV. The increase in scattering between 50 K (diamonds) and 1.56 K (circles) occurs over a broad $l$-range centered around $l$= 0.45. The width of the additional scattering is about twice the width of the additional scattering observed at $E$= 0 meV (errorbars with solid line through the points). The elastic data have been divided by a factor of 5 for plotting clarity.  All lines in both panels are guides to the eye.}
\label{fig1}
\end{figure}
In order to demonstrate the simultaneous presence of frozen as well as dynamic clusters  in Ce(Fe$_{0.76}$Ru$_{0.24}$)$_2$Ge$_2$, we have performed new inelastic neutron scattering experiments using the BT7 triple axis spectrometer at NIST in full polarization mode; we combine these data with previously published data where spectra were collected on the same crystal using the HB3 spectrometer at HFIR, and the DCS spectrometer at NIST.  BT7 was operated \cite{bt7} using neutrons with fixed final energy of 14.7 meV, allowing for the removal of higher order neutrons by means of PG-filters in the scattered beam. $^3$He polarizers were utilized in the incident and scattered beam. The time-dependence of these polarizers was well characterized during the course of the 5 day experiment on BT7, allowing for a full separation between the spin-flip and non-flip signals using the technology and software developed at NIST. All polarized scattering experiments were set up with neutron polarization direction parallel to the direction of momentum transfer so that only the magnetic scattering would show up in the spin-flip channel. We show characteristic data in Fig. \ref{fig1}. The errorbars in all our figures represent plus or minus one standard deviation.\\
The observance of frozen clusters is straightforward, and has already been reported on: previous DCS experiments\cite{wouterprb} showed that the additional scattering that developed below 1 K at the ordering wave vector $Q_{ord}$= (0,0,0.45) was resolution limited, implying that the cluster reorientation frequency had to be longer than 80 ps, therby identifying these clusters as frozen and verifying one part of the prediction of reference [\onlinecite{vojta}].\\
In order to interpret the scattering associated with the reorientations of the superspins of the isolated clusters, we need to demonstrate that the observed quasi-elastic scattering is magnetic in origin; this is shown in Fig. \ref{fig1}. Second, we need to show that the reorientation rate of smaller clusters is higher than that of larger clusters.\\
Our task is facilitated by the fact that once clusters form upon cooling in Ce(Fe$_{0.76}$Ru$_{0.24}$)$_2$Ge$_2$, these clusters persist upon further cooling\cite{wouterprb,heitmannonline}. That is, once a cluster has formed and been forced to order through finite size effects, then this cluster is no longer subject to Kondo screening. As a result, by doing temperature dependent studies, we can first observe the dynamics of the smallest clusters at the higher temperatures. Then, upon further cooling, we observe the dynamics of larger clusters as additional scattering on top of the scattering due to the smaller clusters. We show in Fig. \ref{fig2} that this behavior can also be observed in quasi-elastic scattering by polycrystals.\\
\begin{figure}
\begin{center}
\includegraphics*[viewport=100 110 550 548,width=70mm,clip]{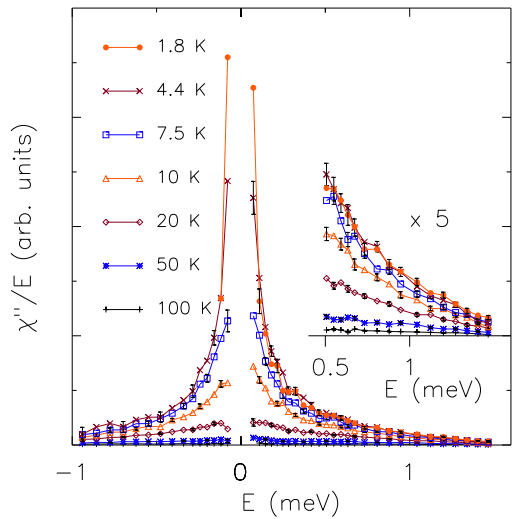}
\end{center}
\caption{Shown are quasi-elastic neutron scattering data on polycrystalline  Ce(Fe$_{0.76}$Ru$_{0.24}$)$_2$Ge$_2$ at $|Q_{ord}|$ = 3 nm$^{-1}$, demonstrating the emergence of magnetic clusters. These data (replotted from reference [\onlinecite{wouterprl}]) were taken on the IN6 spectrometer at the ILL. The energy resolution of the spectrometer is 0.07  meV. Upon cooling from 100 K to 2 K magnetic correlations are seen to develop in the imaginary part of the dynamic susceptibility $\chi$\texttt{"}$(q,E)$ with the values at any given $T$ exceeding those at higher $T$. There is no sign of intensity migrating from the higher energies to lower ones, as emphasized by the tail of the data shown on an expanded vertical scale. Thus, the clusters present at the high $T$ persist down to the lowest $T$, and that the additional scattering is due to newly minted clusters that peel off the infinite cluster upon cooling. Note that the data at 4.4 K and 1.8 K mostly differ in the region $|E| < $ 0.2 meV, implying that very close to the quantum critical point the infinite cluster is rapidly breaking apart, but still spawning fluctuating clusters.  All lines are guides to the eye. For plotting clarity, only one in every three errorbars has been plotted in the main figure, and only every second errorbar in the inset.
}
\label{fig2}
\end{figure}
For emerging clusters that persist down to the lowest $T$ while still fluctuating, we expect to observe the following. On cooling, we should see the scattered intensity increase in the vicinity of $Q_{ord}$, with the intensity at lower $T$ exceeding that at higher$T$ for all energy and momentum transfers in the vicinity of $Q_{ord}$. Thus, any new intensity should appear on top of the intensity already present for the case where smaller clusters survive further cooling and keep their reorientation capabilities. In addition, the new intensity should correspond to larger clusters, which corresponds to a sharper signal in momentum space. This intensity should also have a smaller characteristic energy width, reflecting the lower flipping rate of the larger clusters. We have observed all this.\\
 The data in Fig. \ref{fig2} shows that, on cooling, the additional intensity appears on top of the existing intensity, with a decreasing characteristic energy-width. We have verified this on BT7 for $|E| >$ 1 meV (where resolution effects do not hamper the interpretation) for our single crystal. Since these data do not add to Fig. \ref{fig2}, we show them elsewhere. In addition, it had already been shown that the elastic scattering around $Q_{ord}$ followed a similar pattern\cite{wouterprb}: on cooling, additional scattering was observed\cite{heitmannconf} to appear on top of the intensity already present, and this additional scattering was narrower in momentum transfer, reflecting the formation of larger clusters without affecting the clusters already present.\\ 
The data in Fig. \ref{fig1}b is a good demonstration of the above. When measuring at constant energy transfer ($E$= 0 and 1.25 meV), we see that the data at $E$= 1.25 meV are wider than the elastic data. This is exactly as expected for a system that sees smaller, fast-fluctuating clusters emerge at higher $T$, but predominantly slower, larger ones at low $T$. When this happens, the intensity for $E \gtrsim 1/\Delta t$ (with $\Delta t$ the flipping rate of a large superspin) stops evolving on further cooling and its quasi-eleastic width reflects the width at the temperature below which predominantly larger clusters emerge. E.g., the larger clusters that emerge at the lowest $T$ fluctuate so slowly that they do not add to the intensity at $E$= 1.25 meV. As a result, while the width in momentum transfer narrows upon cooling for the elastic channel, the width at $E$= 1.25 meV remains unchanged, reflecting the (persisting) presence of the smaller clusters. For reference, the width at 1.25 meV corresponds to the elastic width at $\sim$ 4 K\cite{wouterprb}. Note that 'normal' critical scattering does not exhibit this (Fig. \ref{fig1}b) behavior.\\ 
Thus, smaller clusters exhibit a higher reorientation rate than larger clusters, as predicted theoretically\cite{vojta}. However, the superspin of larger clusters can still flip, as witnessed in the temperature dependence of the scattering at lower energy transfers (and clearly visible (for $E>$ 0.1 meV) in the high-resolution polycrystalline data shown in Fig. \ref{fig2}). Combining this with the emergence of resolution-limited magnetic intensity that emerges at the very lowest $T$ on the high-resolution spectrometer DCS\cite{wouterprb}, we conclude that  Ce(Fe$_{0.76}$Ru$_{0.24}$)$_2$Ge$_2$ houses both frozen and dynamic clusters, and that the reorientation rate of the dynamic clusters depends on the size of the magnetic clusters.\\
In summary, we have shown that isolated, ordered clusters emerge, upon cooling, in a strongly doped quantum critical system. These clusters can be frozen as well as dynamic, and the degree of freedom associated with the reorientation of the superspin of the dynamic clusters persists down to the lowest temperatures, providing the system with a readily accessible branch of low-energy excitations. This branch is likely the reason behind the observed but unexplained\cite{si,hertz,millis} $E/T$-scaling.\\


\begin{thebibliography}{10}
\bibitem{stewart} G.R. Stewart, Rev. Mod. Phys. {\bf 73}, 797 (2001); {\bf 78}, 743 (2006).
\bibitem{meigan} M.C. Aronson {\it et al.}, Phys. Rev. Lett. {\bf 75}, 725 (1995).
\bibitem{schroder} A. Schr\"{o}der {\it et al.}, Nature (London) {\bf 407}, 351 (2000).
\bibitem{wouterprl} W. Montfrooij {\it et al.}, Phys. Rev. Lett. {\bf 91}, 087202 (2003).
\bibitem{si} Qimiao Si {\it et al.}, Nature (London) {\bf 413}, 804 (2001).
\bibitem{hertz} John A. Hertz, Phys. Rev. B {\bf 14}, 1165 (1976).
\bibitem{millis} A.J. Millis, Phys. Rev. B {\bf 48}, 7183 (1993).
\bibitem{stauffer} Dietrich Stauffer and Amnon Aharony, {\it Introduction to Percolation Theory} (CRC, Boca Raton, FL, 1994).
\bibitem{heitmannconf} T. Heitmann {\it et al.}, J. of Physics conference series {\bf 391}, 012018 (2012).
\bibitem{wouterprb} W. Montfrooij {\it et al.}, Phys. Rev. B {\bf 76}, 052404 (2007).
\bibitem{gaddy} J. Gaddy {\it et al.}, J. Appl. Phys. {\bf 115}, 17E110 (2014).
\bibitem{heitmannonline} T. Heitmann {\it et al.}, J. of Mod. Phys. {\bf 5}, 649 (2014).
\bibitem{heitmannprb} T. Heitmann {\it et al.}, Phys. Rev. B {\bf 81}, 014411 (2010).
\bibitem{vojta} Jos\'{e} A. Hoyos and Thomas Vojta, Phys. Rev. B {\bf 74}, 140401 (R) (2006).
\bibitem{bt7} W. C. Chen{it et al.}, Physica B 397, 168 (2007).
\end{thebibliography}

\end{document}